# Novel electron and photon recording concepts in noble-liquid detectors

**A. Breskin**
*Dept. of Astrophysics and Particle Physics*
*Weizmann Institute of Science*
*Rehovot, Israel*

*E-mail*: amos.breskin@weizmann.ac.il

ABSTRACT:
We present several novel ionization-electron and scintillation-photon recording concepts in noble-liquid detectors, for future applications in particle and astroparticle physics and in other fields. These involve both single- and dual-phase detector configurations with combined electroluminescence and small charge multiplication in gas and liquid media.

## Contents

## 1. Introduction

For many decades, Noble-liquid detectors have played, a major role in physics experiments and in various other applications. These include particle tracking and calorimetry in particle-physics experiments, neutrinoless ββ-decay, neutrino physics, dark-matter (DM) searches, γ-ray astronomy, medical imaging, contraband detection and more. Detailed surveys of physical processes in noble-liquids, detector techniques, readout concepts and applications are reviewed in [1-7] and in references therein.

Massive Noble-liquid detectors have become the leading instruments in DM searches, and neutrino experiments. They operate either in a single-phase (liquid) [8], or in dual-phase (liquid and gas) [9] configurations. Single-phase detectors are simpler, recording the primary scintillation light emitted promptly at the site of interaction. In addition, ionization charges, deposited amply by relativistic particles (e.g. in neutrino experiments) can be detected and localized, without multiplication, while crossing wire grids within the liquid [10, 11]. A more recent approach is the replacement of the wire grids by perforated electrodes with segmented readout strips [12].

Ionization-electron detection without charge multiplication is not suitable in DM experiments; in the latter, the detection of Weakly Ionizing Massive Particles (WIMPs) is based on the observation of very low energy nuclear recoils originating from their very rare scattering on the noble-liquid nuclei. Dual-phase detectors [9, 7] operate as Time Projection Chambers (TPC), recording two signals: the primary prompt scintillation flash (S1) within the liquid and a secondary delayed signal (S2); S2, generated by radiation-induced ionization electrons liberated at the interaction site, is detected after the electrons' extraction from liquid into the gas phase. S2 signals can be a result of electroluminescence in the saturated vapor phase above the liquid, detected by photo-sensor arrays, or of a multiplied charge in this gas gap. The charge signals can be detected either electronically, e.g. after avalanche multiplication in Large Electron Multipliers, LEM [13] or through VUV photons emitted by the charge avalanche. The photons can be localized using either vacuum photomultipliers (PMT) or more advanced photo-sensor arrays [14-18]; events in detectors emitting sufficient VUV-photon yields can be also recorded with dedicated CCD [19], TimePix3 [20] or other cameras.

While current leading DM experiments employ few-ton liquid xenon (LXe) dual-phase detectors [21, 22, 23], those of the next-generation [24, 25], are designed to reach the 50-ton LXe scale. The next dual-phase liquid argon (LAr) DM experiment, DarkSide [26], targets at the 20-ton scale. The more massive accelerator-based neutrino experiments employ LAr target-detectors. The ~100-ton LAr single-phase, crossed-wires TPC MicroBooNE detector [10] reaches efficient particle identification by combined accurate tracking and calorimetry. The future Deep Underground Neutrino Experiment (DUNE) has investigated large single-phase and dual-phase detector configurations; the experiment recently opted for a ~68-Kton LAr single-phase TPC "far detector" with charge collection on crossed-wires (the status of the 720-ton single-phase LAr ProtoDUNE is given in [11]). There is an ongoing R&D within DUNE on a vertical-drift single-phase TPC with perforated PCB-electrode readout [12].

However, the scaling up of existing techniques, particularly that of dual-phase detectors, may encounter some technical difficulties in deploying large-area wire or mesh charge-extracting electrodes. These mainly relate to keeping constant liquid-to-electrode gap (e.g. might suffer electrostatic electrode-wire sagging and staggering), and avoiding instabilities of the liquid-to-gas interface; both may seriously affect the detector's properties and thus the data quality. Deposited-charge collection in the noble liquid, without multiplication, requires sophisticated low-noise electronics, still setting detection thresholds.

Note that unlike dual-phase TPCs, single-phase detectors permit face-to-face vertical or horizontal detector configurations. The central drift cathode permits applying half of the, somewhat, problematic high voltage across the drift volume. In addition, sporadic-bubble formation in the liquid should be less detrimental in a horizontal geometry.

Thus, in parallel to intensive R&D on novel photo-sensors for future noble-liquid detectors, aiming at the replacement of current PMTs [27], there have been attempts to develop novel concepts of single-phase and dual-phase TPCs; among others, with the aim of overcoming the liquid-to-gas interface issues. The original idea, of ionization-electron multiplication on thin-wires deployed within the noble liquid [28, 29, 30], has lately been undertaken in LXe [31]; the study resulted in very low charge-gain limits, but in electroluminescence photon yields close to 300 photons per electron with 10μm in diameter wires. The authors set the electric-field thresholds for EL (~400KV/cm) and avalanche multiplication (~700KV/cm). The values are of the same order of magnitude as reported in the early works of the "Japanese School" [29] in LXe. A more recent ongoing R&D related to single-phase detectors with amplifying (small avalanche and EL) wire-arrays in LXe is reported in [32]. The authors discuss in detail the general advantages of single-phase detectors for DM searches and provide simulation results of the expected light yields for anode wires located between two mesh electrodes. E.g. for a charge gain of ~3, they expect ~100 photons/$e^-$ with 10 μm anode wires polarized at 5KV in their configuration, and 200 photons/$e^-$ with 50 μm wires, at 14.3 KV. Note however, that at such high potentials, the electrostatic forces may cause wire staggering; this would lead, particularly with long anode wires, to electrical instabilities and to loss of energy resolution. Additional suggestions for single-phase detectors with other wire configurations, e.g. allowing for reduced EL-photon shadowing by the wires, are discussed in [33].

There have been other attempts of charge multiplication in noble liquids: on needles [34, 35] and in a LAr-THGEM (THGEM - Thick Gas Electron Multiplier [36]) [37]; apparently, in both cases, the avalanches might have developed in small gas pockets. A broad discussion on the EL threshold in LAr, for which a large discrepancy exists between experimental (~60kV/cm) and theoretical (~3MV/cm) values, is provided in [5]. A successful attempt of charge multiplication with a Micro-Strip Plate (MSP) immersed in LXe, resulted in a low charge gain of ~10, with electrodes having 8μm wide anode strips [38]. More details on MSP-based multipliers are provided in Par. 3.1.

In this article, we summarize in brief the state-of-the-art of some new, currently investigated, ionization-electron and scintillation-photon recording concepts. Novel directions are proposed for optical recording of interaction-induced EL- and avalanche-photons in single-phase detectors.

## 2. Recent Detector Concepts under evaluation

### 2.1 The Bubble-assisted Liquid Hole Multiplier

In its original configuration, the Liquid Hole Multiplier (LHM) [39] was conceived as a "purely" single-phase detector, sensitive to both ionization electrons and scintillation photons - induced by radiation within the noble liquid. The proposed concept relied on cascaded CsI-coated perforated electrodes immersed in the liquid, designed to amplify the initial charges by EL. Investigations have revealed that its operation relies indeed on EL induced in a gas bubble trapped underneath an immersed perforated electrode, both in LXe [40] and LAr [41]. In its current configuration (Figure 1), the LHM consists of a perforated electrode (e.g. Gaseous Electron Multiplier (GEM) [42] or a THGEM) immersed in the noble liquid, with heating wires generating (initially) a stable vapor bubble underneath.

With the electrode's top surface coated with a CsI (or other) VUV-photocathode, and under proper electric fields, radiation-induced scintillation photons (S1) extract photoelectrons that drift into the electrode's holes; they cross the liquid-vapor interface into the bubble, where they induce an EL signal (S1'). Similarly, radiation-induced ionization electrons in the liquid drift into the electrode's holes and induce a delayed EL signal (S2) across the bubble (Figure 1b). EL-photons recorded with an array of photo-sensors, e.g. Silicon Photomultipliers (SiPM), provide the event's deposited energy and localization. It is expected that compared to a large-size "standard" dual-phase detector, where variations in the EL gap (see above) would affect energy resolution, the vapor bubble is confined in a well-defined geometry and liquid- vapor interface. This "local dual-phase" concept is expected to yield superior results in terms of stability and energy resolution.

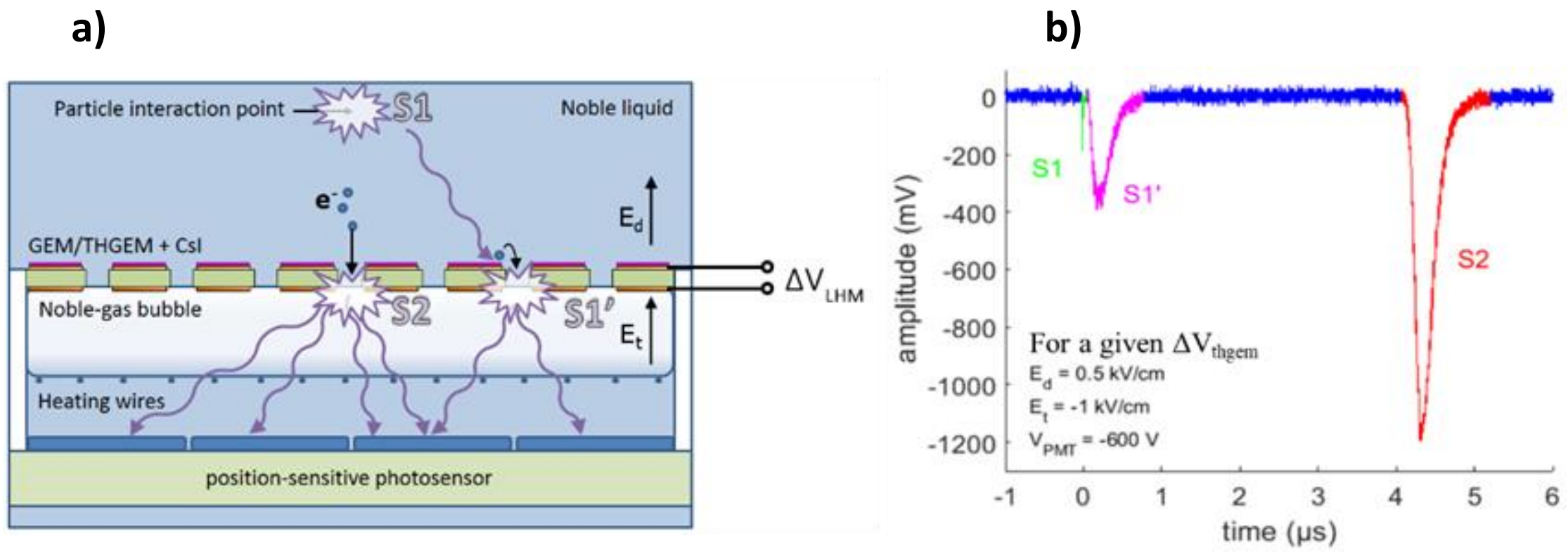

***Figure 1*** *a) Schematics of an LHM detector. A perforated electrode is immersed in the noble liquid, with a potential* $\Delta V_{LHM}$ *applied across. Radiation-induced ionization electrons and scintillation-induced photoelectrons from a CsI VUV-photocathode (deposited on top of the perforated electrode) are focused into the electrode's holes; both cross the liquid-vapor interface, inducing EL signals (S2 and S1', respectively) in the (~1.6mm thick) bubble. The signals are recorded by photo-sensors located underneath. b) Typical alpha-particle induced S1' and S2 signals, detected by a PMT; the prompt weak S1 pulse is induced by scintillation photons viewed directly, through the holes, by the PMT For details see [43].*

delayed EL signal (S2) across the bubble (Figure 1b). EL-photons recorded with an array of photo-sensors, e.g. Silicon Photomultipliers (SiPM), provide the event's deposited energy and localization. It is expected that compared to a large-size "standard" dual-phase detector, where variations in the EL gap (see above) would affect energy resolution, the vapor bubble is confined in a well-defined geometry and liquid- vapor interface. This "local dual-phase" concept is expected to yield superior results in terms of stability and energy resolution.

The LHM concept has been studied extensively with small prototypes with various perforated electrodes, both in high-purity LXe [43] and LAr [44]. Under optimal conditions, the LHM detector yielded ~ 400 photons/$e^-$/4π in LXe. Energy resolutions with 5.5 MeV alpha particles were of the order of 5-6% RMS, for both S1' and S2 signals. The localization resolution of alpha particles was of the order of 0.2mm RMS in LXe [45]. Worse, though very preliminary, results in terms of energy and localization resolutions have so far been obtained in LAr [41].

The photo-detection efficiency (PDE), measured so far in LXe by counting S1' photoelectrons (with reference to a known single-photon source) are of the order of 3-4 % [43], compared to an expected value of ~20%; the latter is derived from the known effective quantum efficiency ($QE_{eff}$) value of CsI in LXe [44, 46, 47] under the current experimental conditions. The expected $QE_{eff}$ values in LXe vs the applied potential, are plotted in Figure 2, for THGEM, GEM and single-conical GEM (SC-GEM) LHM electrodes.

The very low measured PDE values suggest electron losses induced, on the one hand, by somewhat low photoelectron collection efficiency into the holes; our current hypothesis though, focuses on an inefficient transmission of electrons through the liquid-gas interface into the bubble. It might be affected by the location and shape of the liquid-to-gas interface at the hole's bottom. Possible hypotheses regarding the physics processes at the origin of this problem are discussed in [43, 47] (and in Par. 4). Such electron losses could be also at the origin of the relatively low energy resolution quoted above for alpha particles, compared to the expected one, e.g. for S1, of ~0.7% RMS from pure photon statistics and the measured one by XENON, of ~1.5% RMS [48]. The results of a recent study related to the liquid-gas interface shape and its dependence on the electric fields are discussed in [49]. Model simulations discussed in this work could permit, in principle, designing electrode layouts with optimized electric fields at the interface - to enhance electron transmission, thus the PDE and energy resolution.

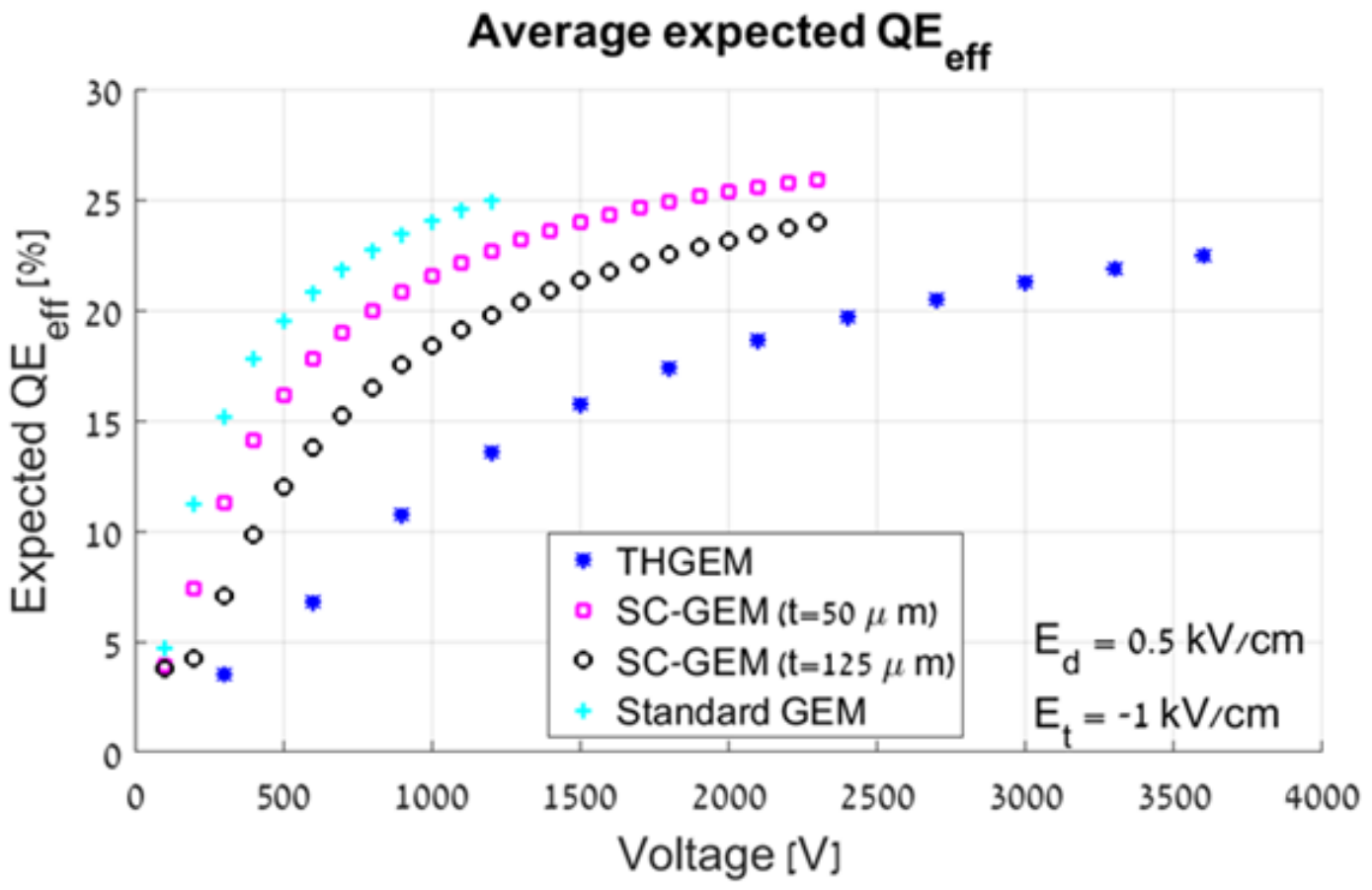

*Figure 2. Expected average $QE_{eff}$ across the entire surface of a perforated electrode immersed in LXe, as a function of voltage across the electrode. They were computed (using COMSOL®) for different LHM electrodes; electric field values: $E_d$=0.5 kV/cm and reversed transfer field, $E_t$=-1 kV/cm. (from [47]).*

A considerable enhancement of the currently-reached PDE value of the LXe-LHM (e.g. >15%) and a proof of long-term stability of large multiplier modules, are the prerequisites for conceiving large-volume "local dual-phase" LHM-based TPCs. A potential scheme, proposed within R&D program of the future DARWIN DM experiment [24], is shown in Figure 3. It consists of covering the TPC bottom with LHM modules, of the kind shown in Figure 1.

### 2.2. Two-phase TPC with immersed perforated electrode

The electron collection and transmission losses through the liquid-gas interface in the bubble-assisted LHM, discussed above, and on the other hand, the high QEeff values of CsI photocathodes in LXe [46, 44], as compared to vacuum, have led to a new LHM concept presented in [50]. The idea (Figure 4) consists of a CsI-coated perforated electrode (on its bottom face), immersed in the noble liquid, e.g. a Liquid-THGEM (L-THGEM), with a second one, e.g. a Gas-THGEM (G-THGEM), located above in the gas phase. It is known that a single-THGEM yielded a charge gain ~200 in dual-phase Ar [51]; a single-GEM yielded a charge gain ~150 in dual-phase

Xe [52]. The operation of such single- and cascaded Cryogenic Avalanche Detectors (CRAD) is reviewed in [4] with numerous references therein.

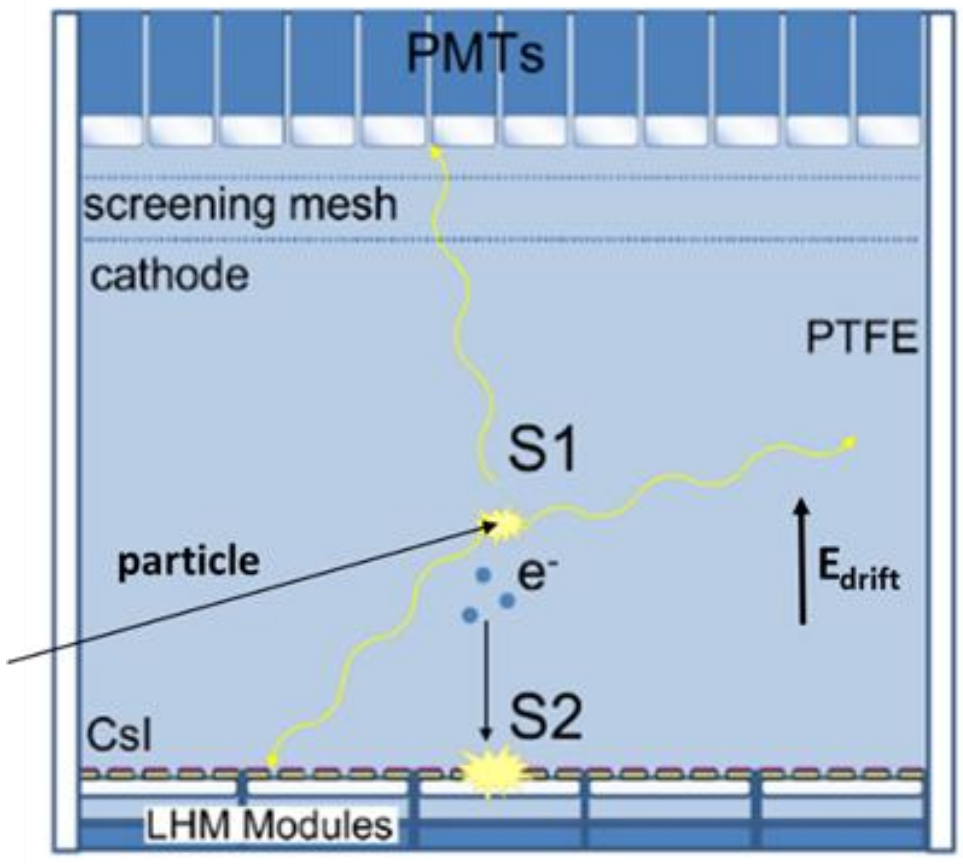

***Figure 3*** *Schematic design of a liquid-only single-phase TPC with bubble-assisted LHM modules at the bottom. They detect S1 signals of a fraction of the scintillation photons and S2 signals induced by ionization electrons drifting down towards the LHM holes. Accurate S2-based position reconstruction is provided by an array of photo-sensors (e.g. SiPM) located below the LHM modules. The top PMTs can be optionally replaced by other sensors or by a reflective surface.*

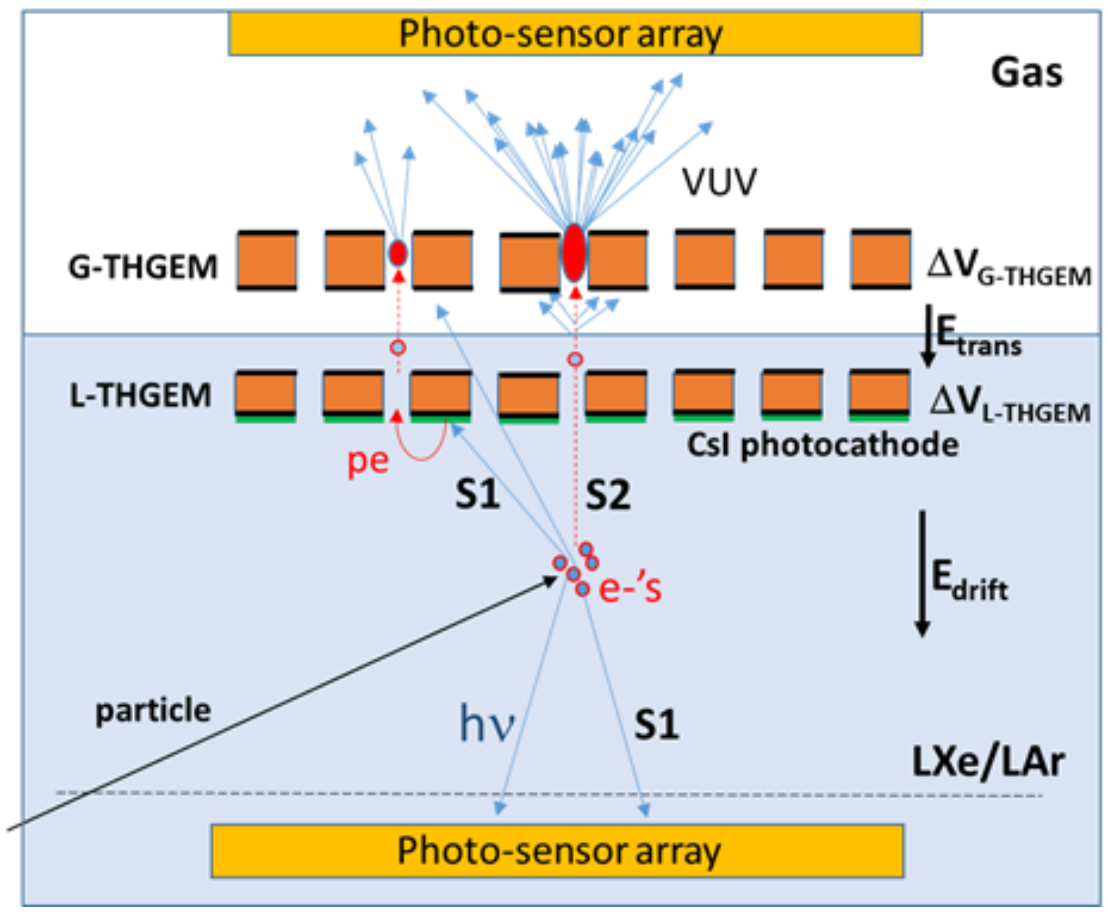

***Figure 4*** *Schematic design of a two-phase TPC with immersed CsI-coated perforated electrode (here a THGEM) and a second one in the vapor phase. In the shown scheme, scintillation S1 photoelectrons from CsI and ionization electrons are collected into the L-THGEM holes. They are efficiently transferred into the vapor, where they induce fast EL and avalanche photons in a second (here a G-THGEM) element; these are detected and localized by photo-sensors located above the G-THGEM. A photo-sensor array or a reflective cathode (not shown) at the bottom, would enhance the detectable S1-photon yield. The top L-THGEM electrode replaces the usual Gate grid of dual-phase detector, which simplifies the construction.*

In our concept, the EL and avalanche photons are viewed by a photosensor array deployed above the gas multiplier (a G-THGEM in Figure 4). The liquid-vapor interface is located above the immersed perforated electrode (here a L-THGEM), CsI-coated at its bottom face - with no bubble underneath (unlike in the original LHM concept [39]).

Ionization electrons deposited within the drift volume below the electrode and S1-induced photoelectrons emitted from the photocathode are focused into the L-THGEM electrode holes, traversing them without multiplication, with almost no losses [50] (as shown below). An intense transfer field ($E_{trans}$) above the immersed electrode (taking here the role of the "Gate" in conventional dual-phase TPCs) ensures efficient transmission of the electrons into the vapor phase; they are focused into a second perforated electrode, e.g. a G-THGEM (in Figure 4), where they induce intense S1 and S2 VUV-photon flashes from combined EL and charge-avalanche in the holes. Other electrode types can be used either in the liquid or in the vapor phase; in liquid, to optimize photoelectron emission from CsI ($QE_{eff}$) (Figure 2) as well as electron and photoelectron transmission through the holes; in gas phase, to reach higher photon emission yields towards the photo-sensors.

Among other possible perforated electrodes, in the liquid, are "standard" bi-conical GEM or single-conical GEM, discussed in [42,43]. In the gas phase, a good candidate could also be a Micro Hole & Strip Plate (MHSP) [52]. It provides a two-stage avalanche: electrons entering the holes undergo a first avalanche step, whose electron cloud is focused onto thin strips patterned on the electrode's top face. The MHSP was reported to yield (by combined EL + charge avalanche) ~$7x10^4$ photons/electron in Xe gas at room temperature – about an order of magnitude larger than of a standard GEM (in the same setup) [54]. A similar, thicker MHSP-like multiplier could be the more robust Thick-COBRA (THCOBRA) [55]. It should be mentioned that another recent concept proposes very-thick GEM structures made on wavelength-shifting transparent polymers (e.g. polyethylene naphthalate (PEN)) for primary- and EL-photon optical recording in dual-phase TPCs [56].

The different perforated electrodes discussed in this work are depicted in Figure 5. Their geometrical and electrical parameters should be carefully adjusted to maximize the photon yields. The main advantages of a potential dual-phase TPC based on this concept are that single-VUV-photon detection efficiencies (PDE) can reach values of ~ 20% [44, 47], and that individual VUV photons generate large-amplitude fast signals - that cannot be misinterpreted by the photosensors' dark counts. This allows the use of SiPM [10], CMOS-SPAD [57] and other lower-cost digital sensors, despite their relatively high dark-count rates. Located in the gas phase, these sensors should be either sensitive to VUV or coated with a wavelength shifter (WLS); e.g. Tetraphenyl Butadiene (TPB) is generally employed in LAr [1].

The efficient collection of photoelectrons from the bottom L-THGEM (Figure 4) and of electrons drifting in the liquid towards the L-THGEM, as well as their transfer towards the liquid-vapor interface were recently demonstrated [50]. The preliminary results are shown in Figure 6. A saturation is observed of the normalized currents collected on a mesh placed in the liquid phase above the L-THGEM (not shown in Figure 4); they originate from the CsI-emitted photoelectrons and from UV-induced ones (on a photocathode, mimmicking ionization electrons) drifting from the liquid underneath the L-THGEM - indicating a very high collection and transmission efficiency of the initial charges. Transfer-field values >5 kV and >3kV/cm should assure full electron transmission from LXe and LAr, respectively, into the gas phase [58, 3]. The proof-of-concept in a dual-phase mode, with various electrode configurations, and the assesement of the detector's properties, are the subject of current studies.

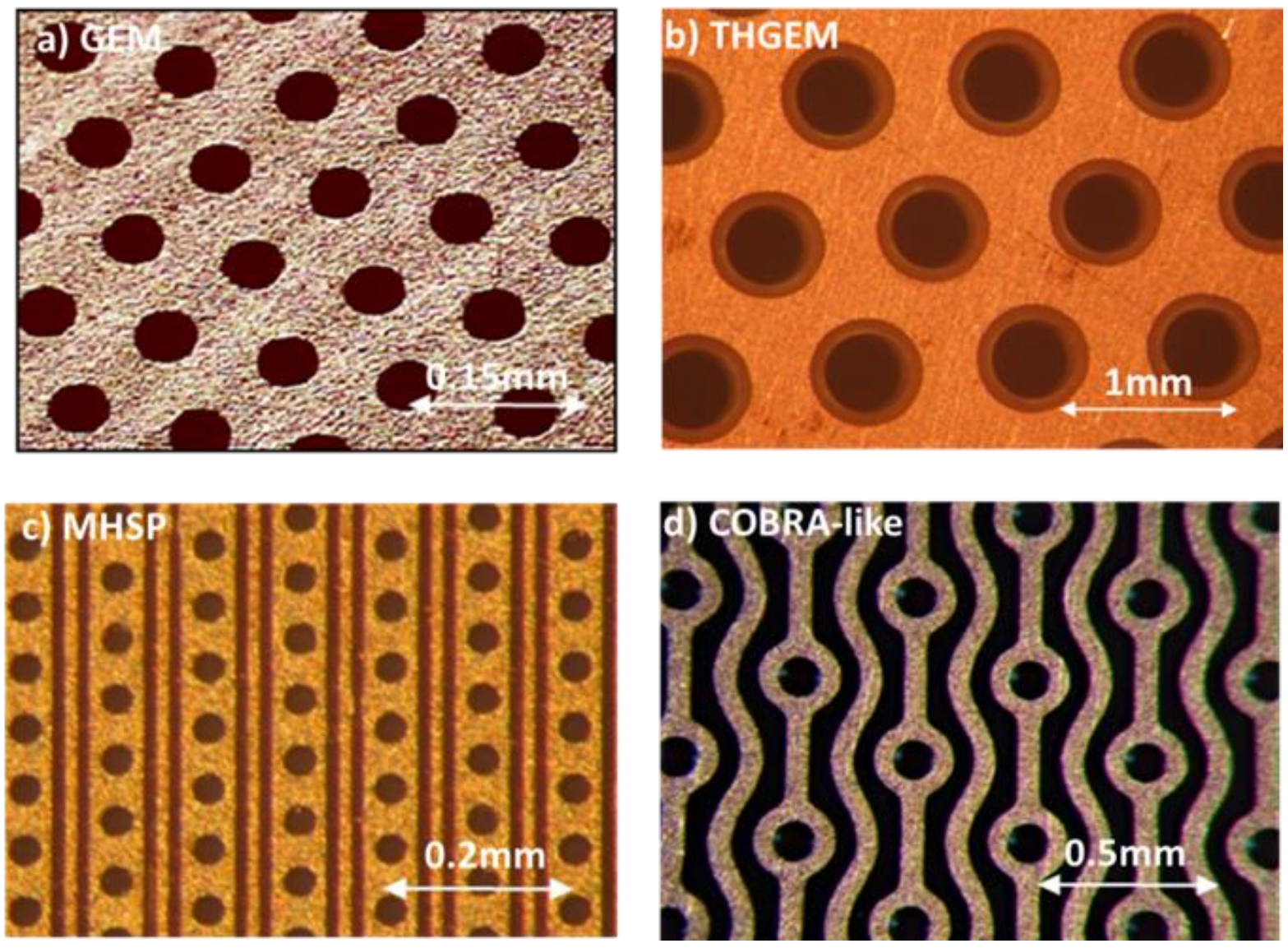

*Figure 5. Different perforated electrodes discussed in this work. a) GEM, with typically ~60μm in diameter holes etched in 50 μm thick Kapton foil; b) THGEM, with typically ~0.5mm in diameter holes drilled in sub-mm thick FR4 plate and with etched hole-rims to prevent discharges; c) MHSP, with ~60μm holes and anode/cathode strips etched in 50μm thick Kapton; d) COBRA-like electrode, here a 125-COBRA, with Cu/Kapton: 120μm holes and anode/cathode strips etched in 125μm thick Kapton. A THCOBRA has a similar pattern, but with typically ~0.3mm in diameter holes drilled in 0.4mm thick FR4 plate. These electrode parameters vary according to their respective applications.*

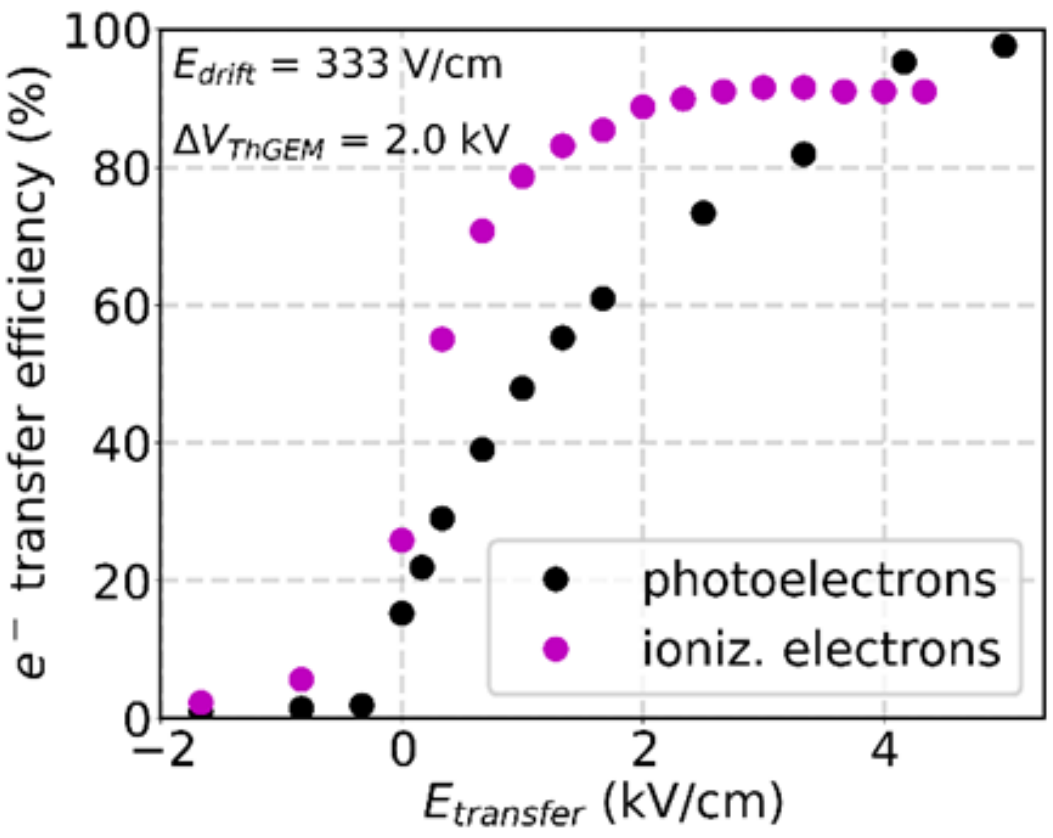

***Figure 6*** *Transfer efficiency (preliminary) of ionization electrons and photoelectrons through L-THGEM holes, vs the transfer field $E_{trans}$ between the L-THGEM and a collection mesh located within the LXe, above it, in the configuration depicted in Figure 4 (the mesh is not shown in the figure). The currents were induced by electrons drifting in the liquid (S2) and those induced by photoelectrons emitted from the photocathode (S1). The collected-current values at the mesh were normalized to the initial electron and photoelectron ones. THGEM geometry: 0.3 mm hole diameter, 0.7mm pitch, thickness 0.4mm; $E_{drift}$=330 V/cm; Voltage across the L-THGEM $\Delta V_{L\text{-}THGEM}$=2kV. Data from [50].*

## 3. Novel single-phase detector ideas

In this paragraph, we present some new ideas of ionization-electron and VUV-photon detection concepts in single-phase noble liquid detectors. Though some of them might seem rather exotic, they are presented here to trigger an interest among the younger generation of researchers in Detector Physics.

### 3.1 Single-phase detectors with Micro Strip Plate (MSP) multipliers

The proposed concept consists of detecting ionization electrons (S2) in a noble liquid via VUV photons emitted by EL and optionally additional small charge avalanche, at the vicinity of narrow anode strips of a Micro Strip Plate (MSP) deposited on a VUV-transparent substrate immersed in the liquid. The primary S1 VUV-scintillation photons can be detected by photosensors, in different configurations, as discussed below. The operation principle, in which anode-wire arrays [32] are replaced by anode strips printed on a transparent substate, is shown in Figure 7.

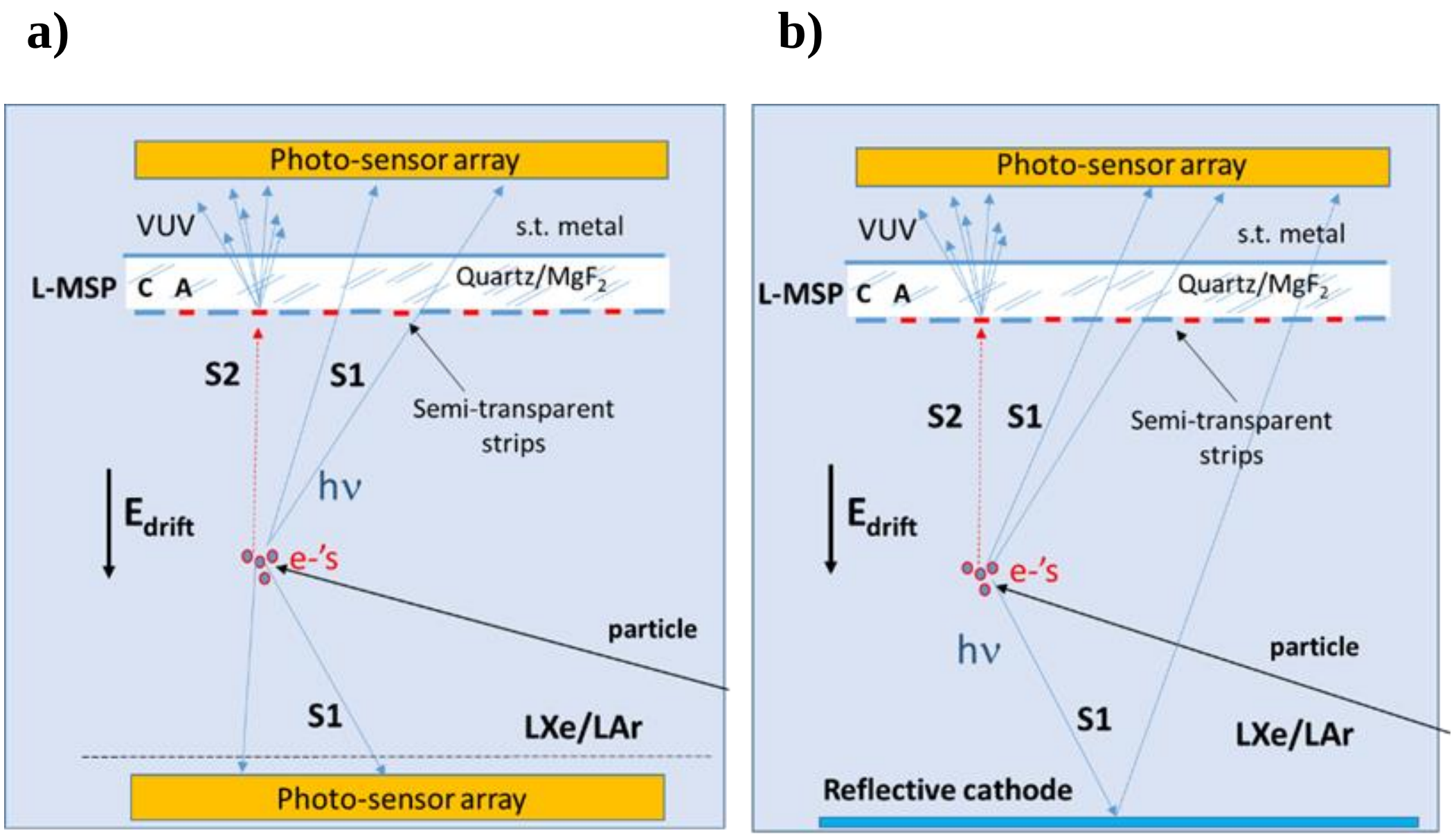

*Figure 7. Two configurations of a single-phase TPC with a L-MSP electrode (here an MSGC-like MSP) with semi-transparent (few-nm thin Cr or Ni) anode (A) and cathode (C) strip electrodes patterned on a VUV-transparent substrate (and an optional few-nm thin Ni or Cr back-plane). Ionization electrons induce VUV photons by EL + small avalanche near the anode strips; they are detected, as well as a fraction of S1 scintillation photons, through the substrate, by the top photo-sensors. Another fraction of S1 photons are either detected by bottom photo-sensors (a), or reflected by a mirror-cathode upwards to the top photo-sensor (b).*

The original MSP concept, that of the Microstrip Gas Chamber MSGC, was introduced by Oed [59] with the idea of reproducing an electric-field configuration similar to that of Charpak's Multiwire Proportional Chamber (MWPC [60]), on a much smaller geometrical scale – so as to enhance radiation-localization resolution. In its original configuration, the MSGC electrode (Figure 8 left) consists of a sequence of alternating thin conductive anode and cathode strips deposited on an insulating support - usually conductive glass, but also on other substrates [61]. The application of appropriate potentials to the anode- and cathode-strips, sets an electric field configuration (Figure 8 right) leading to large proportional charge-avalanche multiplication (~$10^4$) at atmospheric pressure, in selected counting gases. Avalanche ions are collected by the neigbouring cathode strips. In noble gases, EL occurs prior to the onset of avalanche multiplication. To reach a few tens of μm resolutions, the MSP cathode strips widths are typically 30 - 60 μm; anode strips are 5 - 10 μm thin, at about 200μm pitch. A conductive film on the electrode's back surface prevents charging up of the insulator substrate between strips, by repelling avalanche-induced positive ions off the surface. Such potential charging up is not expected under the proposed conditions of EL in liquid with an optional small (<10) charge gain. The active area of MSGC detectors reaches 240x240 $mm^2$ [62]. For a MSP review see [61].

While many works describe the operation of MSP-MSGC electrodes in various gas mixtures, to the best of our knowledge, only a single work relates to its detailed studies in a noble liquid (LXe) [38] (some unsuccessful trials were reported in LAr in [35]). The authors of [38] measured with the LXe MSP (L-MSP) a small charge multiplication, of about 10, in LXe; they did not provide data on the VUV photoyield induced at the strip vicinity by EL or combined EL and avalanche. Like in a counting gas, the photoyield in a noble liquid should depend on the electric-field configuration, namely the MSP electrode geometry (e.g. anode- and cathode-strip width and pitch) and on the applied potentials.

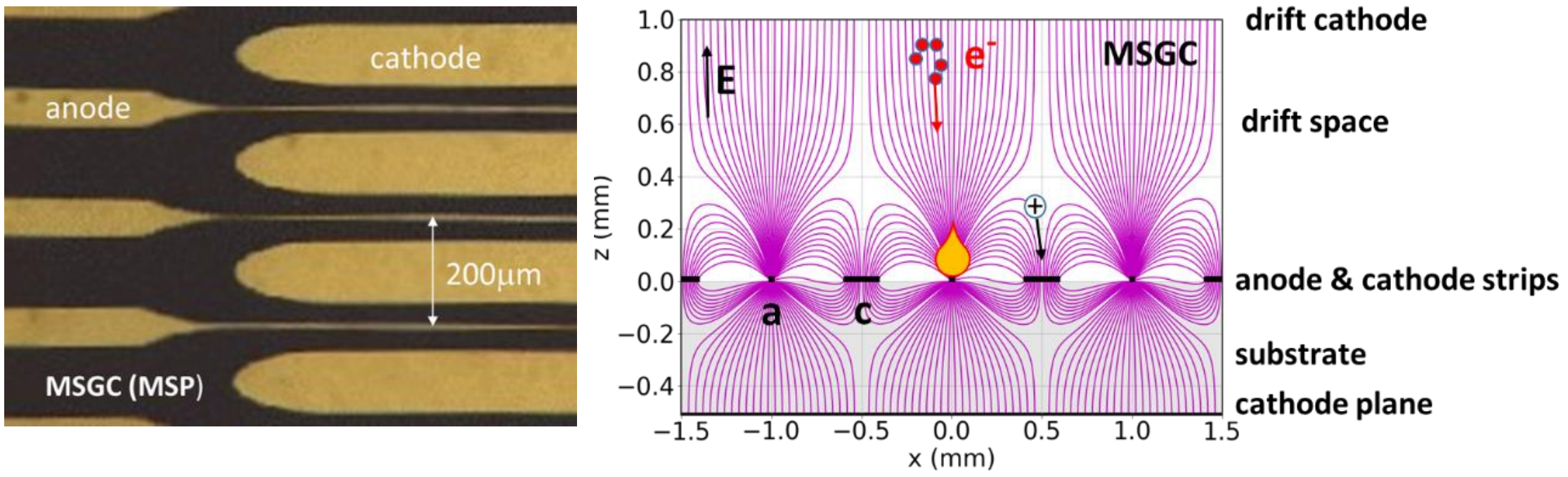

*Figure 8. Left: A photograph of a typical MSGC (MSP) electrode. Alternating thin conductive anode and cathode strips (here with 200μm pitch) are deposited on an insulating support (usually glass), with an optional backplane-cathode. Right: Field-line distributions (COMSOL) in a different electrodes' geometry. Charges deposited in gas, drift along the field lines to the anode strips, where avalanche (shown) occurs; in noble gases EL (not shown) occurs as well. The field value and shape near the anode strips, is set by the potentials applied to the anode ($V_a$), cathode ($V_c$), cathode backplane ($V_b$) and drift ($V_d$) electrodes. In this example, estimated for operation LXe: substrate=0.5mm; anode-strip width=5μm; cathode-strip width=200μm; drift-gap=1.9mm; strip pitch=1mm. $V_a$=5KV; $V_c$=0; $V_b$=0; $V_d$=-300V.*

Besides the regular MSP (MSGC) electrode configuration [59], other types were proposed by several groups (as reviewed in [61]), with the aim of protecting the electrodes from discharge-induced damage to the few-μm thin anode strips. Among them, the Microgap Chamber [63], that like the MSP [38] was investigated in LXe [64]; no charge gain was obvserved but simulations predicted charge gains above 10 in some conditions. Two other microstructures, relevant to this concept, were proposed and investigated in standard counting gases at room temperature: the Coated Cathode Conductive Layer electrode (COCA COLA) [65] and the Virtual Cathode

Chamber (VCC) [66]. They are depicted in Figure 9, with their respective electric field-line configurations (for similar parameters to that of the MSGC of Figure 8).
Unlike the MSGC, the COCA COLA configuration has thin anode strips on one face with neighboring cathode strips located on the other face of the insulator. The VCC has an array of thin anode strips on one face and a plain cathode film on the other one. Note that in "standard" tracking applications, the insulator should be preferably of relatively low resistivity, or coated with an appropriate low-resistivity material, e.g. diamond film [67]. E.g. in MSGC-like structures it avoids instabilities due to inter-strip (anode-cathode) charging up by avalanche ions at large avalanche gain and high-rate operation. Note that none of these applies to the proposed concept (e.g. potentially very low charge gain or none) and to its potential applications in DM, neutrino physics and oother rare-event experiments.

The calculated (with COMSOL) electric-field distributions vs the distance Δr from the anode-strip surface (along the z-coordinate shown in Figures 8, 9) are depicted in Figure 10 for the three strip-electrode configurations discussed above.

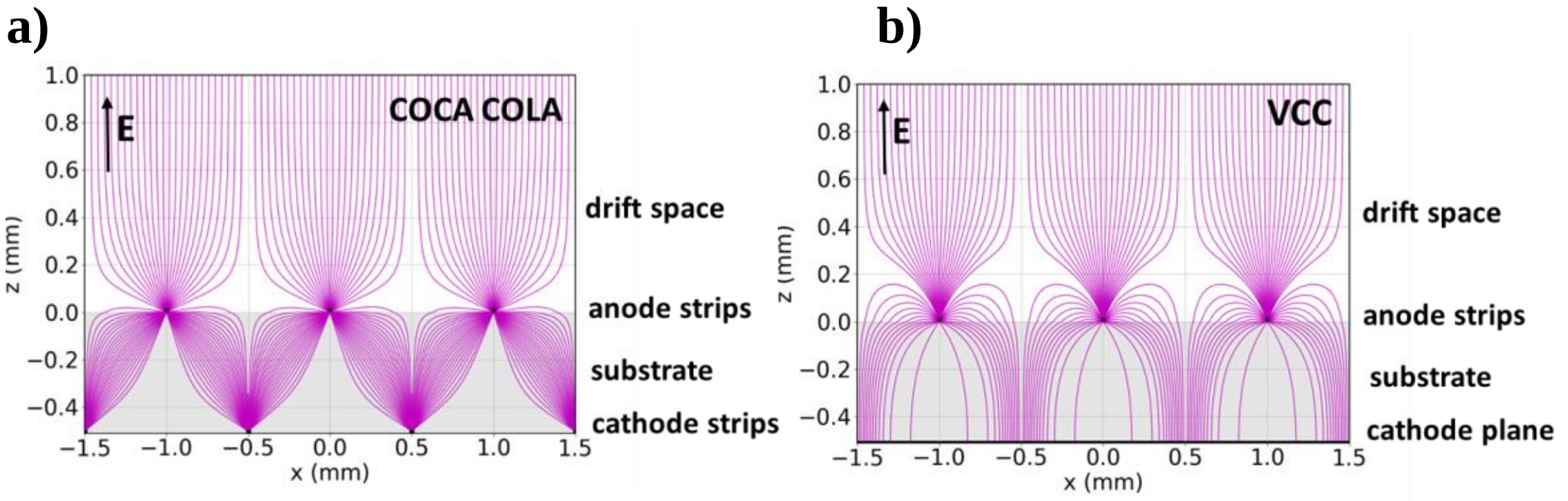

*Figure 9. Schematic views with the field-line distributions in the a) COCA COLA and b) VCC electrodes. Both multipliers have anode strips running on the top face, insulated from the bottom face by the substrate. The COCA COLA electrode has cathode strips, and the VCC a full conductive film - at the back plane. Similarly to the MSGC (of Figure 8), charges deposited in gas, in the drift space, undergo avalanche multiplication at the high electric field in the vicinity of the anode strips. Its value and shape are dictated by the potentials applied to the anode-strip ($V_a$), cathode-strip ($V_c$), cathode-backplane ($V_b$) and drift ($V_d$) electrodes. Simulation parameters (similar the MSGC ones in Figure 8, estimated for LXe): substrate thickness=0.5mm; anode-strip width=5μm; cathode-strip width=200μm (in COCA COLA configuration); drift-gap=1.9mm; strip pitch=1mm. Potentials: $V_a$=5KV, $V_c$=0, $V_b$=0 and $V_d$=-300V.*

The distributions are provided for anode-strips widths of 5 to 50 μm. Note that similar potential values were applied to the electrodes in the three cases, to facilitate comparison. Naturally, the electrode geometries and potentials have to be optimized, in each configuration, for providing the highest-possible light yields (EL or EL+avalanche photons). The EL and charge-multiplication (CM) thresholds indicated in Figure 10 for LXe, were taken from [31]. In a first approximation, one can notice the rather similar electric-field behavior in the MSGC and VCC at the anode-strip vicinity; e.g. for 10mm anode-strip width, the EL threshold starts at Δr~20μm and the CM one at ~10μm; the COCA COLA strip-plate shows lower fields values per given Δr.

The concept proposed here (Figure 7) is to immerse an L-MSP electrode (e.g. MSGC, COCA COLA, VCC or other) in the noble liquid, with strips deposited on a VUV-transparent substrate: e.g. quartz (transmitting Xe 178nm VUV photons) or $MgF_2$ (transmitting both Xe and Ar 128nm light). Note that doping LAr with Xe [5, 68] would effectively shifit the scintillation- and EL-photon wavelength to that of Xe, with the option of using quartz substrates also in LAr. According to Figure 10, the MSGC and VCC electrodes, ptovide similar fields at the strip vicinity. However, the VCC, having anode strips separated by the insulating substrate from the back-plance cathode, should provide more stable operation at higher anode potentials.

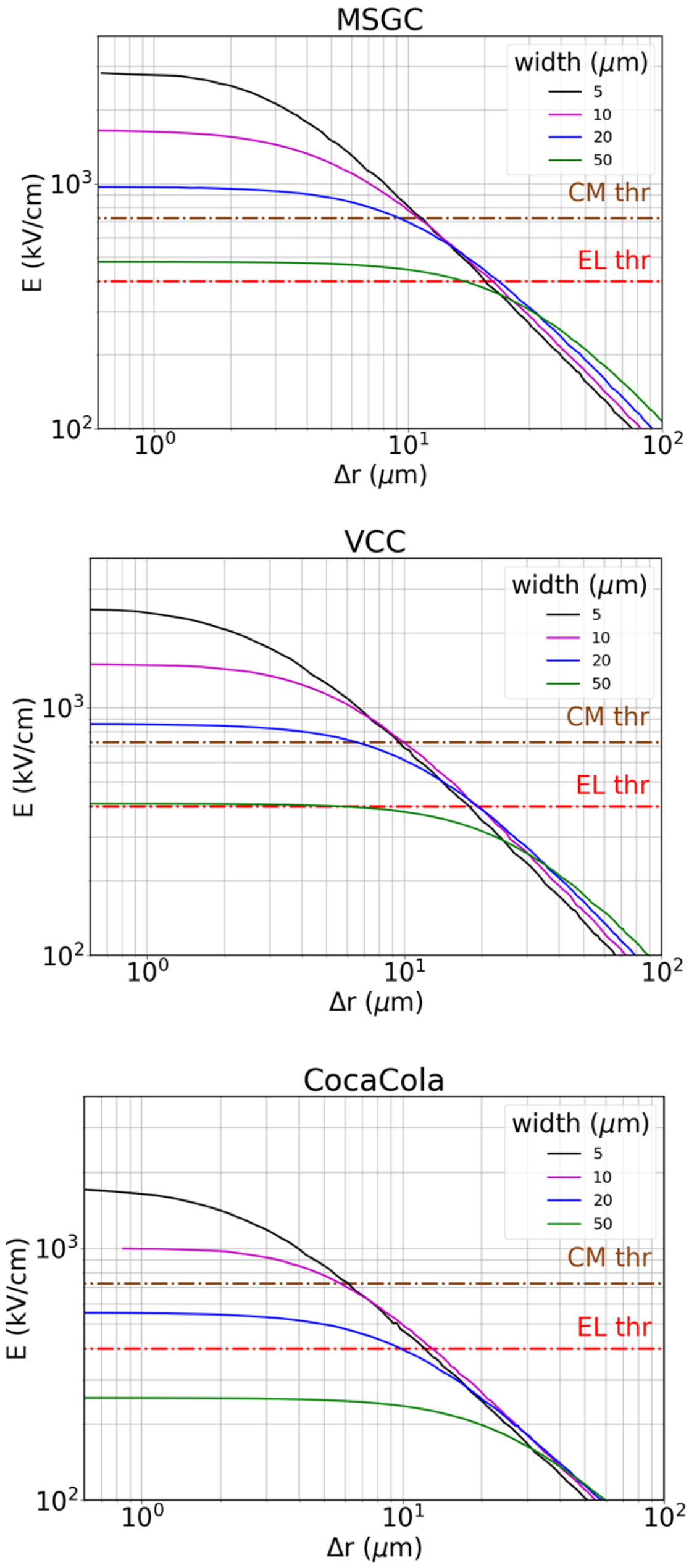

*Figure 10. Simiulated electric-field distributions, vs the distance from the anode-strip surface Δr, in the MSGC, VCC and COCA COLA configurations (depicted in Figures 8 and 9). Similar potentials were applied in the three configurations to the different electrodes. The field value at given Δr, is dictated by the potentials applied to the electrodes. The thresolds of EL and CM (charge multiplication) in LXe are shown as well. Simulation parameters: substrate thickness=0.5mm; anode-strip widths: 5, 10, 20, 50μm; cathode-strip width=200μm (in MSGC & COCA COLA; drift-gap=1.9mm; strip pitch=1mm. Potentials: Va=5KV, Vc=0, Vb=0 and Vd=-300V.*

Note that the idea of replacing the MSGC by a VCC electrode in LXe was also proposed at the end of the nineties by the Policarpo's Group [69]. Their calculated multiplication region extends over several μm above the anode strip, with estimated charge multplication of ~100 at about -2000 V applied to the bottom cathode and to drift electrode. As shown above (Figure 10) our simulated electric-field distributions in a VCC, suggest CM and EL respective thresholds at ~10μm and ~20μm from the strip surface (at 5KV potential difference between the anode strips and the cathode back-plane and -300V on the drift cathode located at 2mm above the strip surface). Higher anode-strip potentials in LXe seem feasible; these should extend the high-field region, thus enhancing the photoyields.

In the two single-phase MSP configurations depicted in Figure 7 (with MSGC-like L-MSP), a fraction of VUV photons induced in the liquid by primary scintillation (S1) and those induced at the anode-strip vicinity by EL (and small avalanche) are transmitted through the substrate to photo-sensors located in the liquid above the L-MSP. As with thin immersed wires [31], a fraction (configuration-dependent) of photons would be masked by the strips. This "lost" fraction depends on the strip geometry and on the applied potential – namely on the distance from the strip surface ($\Delta r$ in Figure 10), where the field reaches the EL and the avalanche thresholds. With distant anode and cathode strips (e.g. COCA COLA and VCC), manufactured with few-nm thin Cr, Ni or other UV-transmitting conductive coatings, part of the "normally masked" VUV photons will still reach the photosensor. E.g. the respective transmission values of ~180nm VUV photons (of Xe or Xe-doped LAr) through 2nm thin Ni and Cr films are 80% and 70%; they are reduced to 40% and 30% for 10nm thin Ni and Cr, respectively [70]. Another fraction of the S1 photons is detected with photo-sensors located at the bottom (Figure 7a) or reflected by a mirror-cathode located at the bottom (Figure 7b).

### 3.2 Single-phase detectors with cascaded perforated electrode and MSP

In a more advanced configuration, depicted in Figure 11, a perforated electrode (e.g. L-THGEM), CsI-coated at its bottom and optionally with a reflective surface or a wavelength shifter at its top surface, is inserted below an L-MSP. Ionization electrons and scintillation-emitted photoelectrons from CsI, are efficiently transferred through the L-THGEM holes (see 2.2 above) onto the L-MSP anode strips (deposited on a VUV-transparent substrate). Anode-emitted EL- and avalanche-photons, (and optionally, those reflected from the perforated-electrode top surface) are transmitted to the photosensors.

Note however, that the reflector will smear out the VUV-photon cloud, thus affecting the localization resolution. Furthermore, in case of difficulties in manufacturing and operating an MSP on quartz or $MgF_2$ substrates, the reflective surface on top of the L-THGEM can be replaced by a WLS emitting VUV-induced visible-range photons; however, it would also cause smearing the photon cloud. In such a configuration, the L-MSP substrate can be made of glass, with the advantage of a larger (and probably more economical) selection of visible-sensitive photo-sensors. Note however, that the stability of current WLS materials (e.g. Tetraphenyl Butadiene - TPB) is questionable, both in LAr [71] and in LXe [72]. Another possibility would be to manufacture the MSP patterns on a transparent WLS polymer (e.g. polyethylene naphthalate (PEN)) discussed in [56]; the stability of this or other WLS-polymers in noble liquids requires detailed studies.

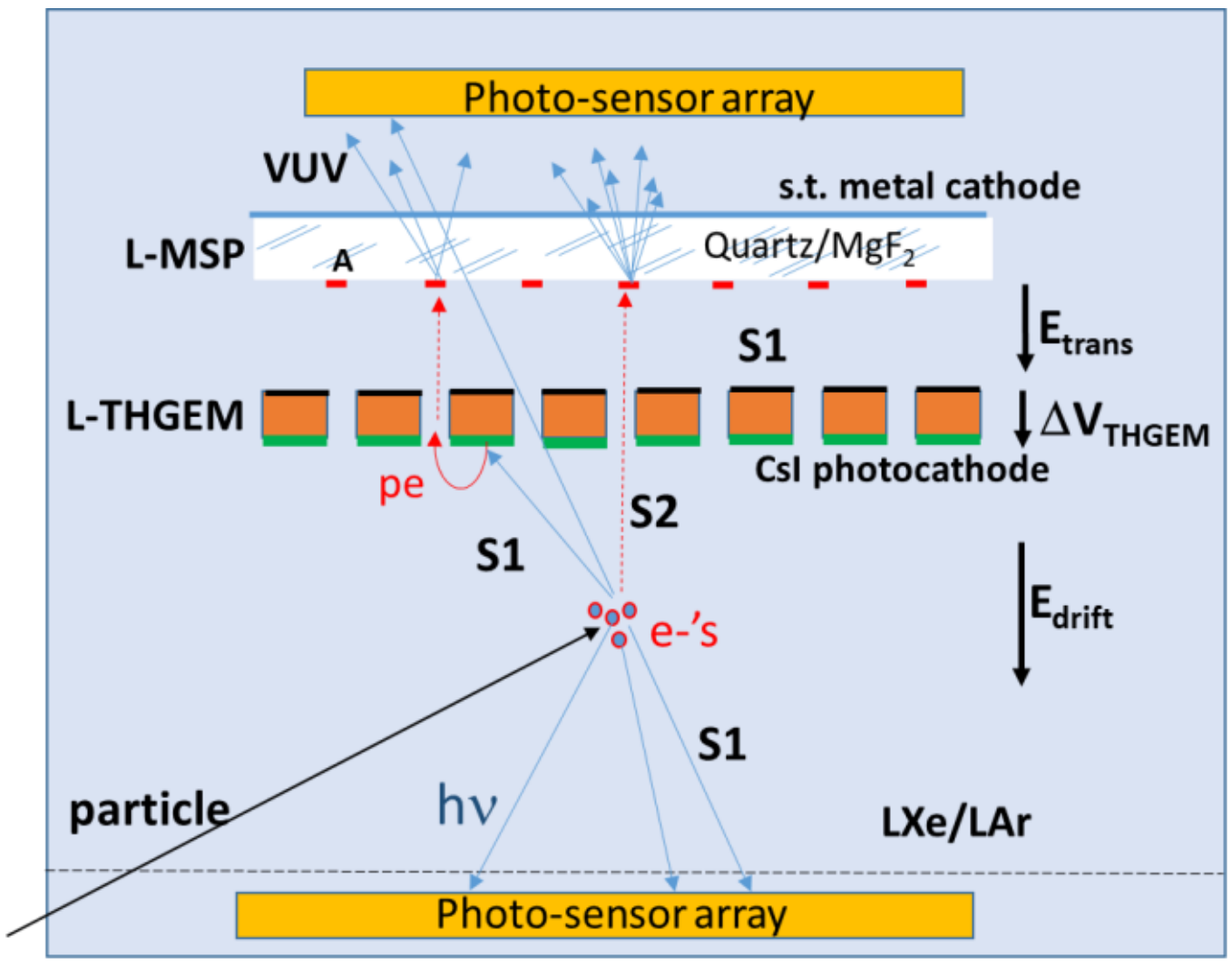

*Figure 11. A single-phase two-stage TPC with a L-THEM coated underneath with a CsI photocathode followed (here) by a L-VCC microstrip multiplier; the L-VCC has semi-transparent (few-nm thin Cr or Ni) electrodes (anode strips A and cathode plane C) patterned on a VUV-transparent substrate. Both ionization electrons and UV-induced photoelectrons from CsI are efficiently collected into the L-THGEM holes; they are transferred to the L-VCC anode strips. VUV photons emitted by EL + small avalanche near the anode strips, are detected (as well as a very small S1 photon fraction traversing the L-THGEM holes) through the substrate, by the top photo-sensors. Another fraction of S1 photons is detected by bottom photo-sensors (as shown), or reflected by a mirror-cathode upwards the photocathode (like in Figure 7b). Optionally, the top surface of the L-THGEM can be either reflective or coated with a wavelength shifter (not shown, see text). An optional wire mesh may be set under the L-THGEM, to enhance the pe extraction efficiency from CsI.*

In the configurations discussed above, the bottom part of the TPC can be covered with photosensors or with a reflective cathode - to enhance the detection efficiency of S1 scintillation photons. A concept incorporating, instead, a CsI photocathode at the bottom would be less attractive. It was proposed by the authors of [73], in a dual-phase LXe TPC concept with a multi-wire electrode in the gas phase; the emitted EL photons being recorded by top PMTs. Note that in such configuration, beside some photon feedback expected from the bottom photocathode, a more significant problem would be the very low $Q_{eff}$ of the CsI photocathode - dictated by the electric field value at its vicinity. E.g. the expected $Q_{eff}$ value in LXe (at 178 nm) is of the order of a few % at 0.3kV/cm TPC drift field [44] (e.g. the field in the LZ experiment [22]).

### 3.3. Single-phase detectors with Micro Hole & Strip Plate (MHSP) – like multipliers

A more elegant and effective configuration is that based on the Micro Hole & Strip Plate, already mentioned in 2.2 above. The MHSP [53] is a two-stage gas-avalanche multiplier that consists of a perforated electrode similar to GEM, but with a top surface patterned into anode and cathode strips, similar to that of a MSP. The MHSP substrate is usually made of a 50 µm thin Polyimide film, Cu-clad on both faces; it is produced in the double-conical GEM technology, with typically 40µm diameter holes in the Kapton and 70 µm diameter in the copper clad, 20µm and 100µm wide anode and cathode strips, at a 200µm pitch (Figure 5c) [74].

In a gas medium, charges drift into the holes, inducing a first charge-avalanche multiplication; the avalanche electrons are collected and multiplied by the thin anode strips patterned on the top surface. As discussed in [53, 75], VUV-photon-induced photoelectrons from a semi-transparent photocathode located above or from a reflective one coating the MSP bottom face, can be detected in the same way. The original gaseous MHSP-based photosensor concept is depicted in Figure 12.

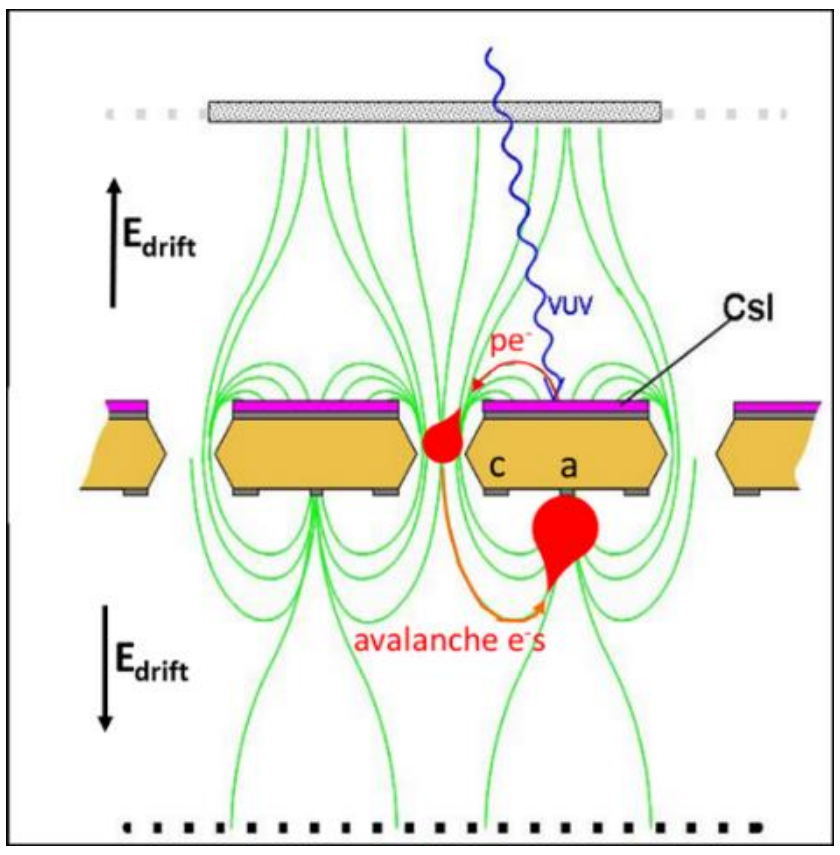

*Figure 12. The UV-MHSP photosensor operation principle in gas. VUV photons interact with a CsI photocathode deposited on the back-plane of the MHSP electrode. Its other face has parallel anode (a) and cathode (c) strips (see Figure 5c). The resulting photoelectrons undergo avalanche preamplification in the holes; the resulting electron swarm is further multiplied under the high field at the anode strips vicinity. Avalanche ions are collected by neighboring cathode strips. The final avalanche charge is detected electronically, or with photosensors recording avalanche- and EL-induced photons (not shown in the figure). The MHSP multiplication process is similar for ionization electrons (not shown), collected from the drift volume.*

Other top-surface patterns around the holes were investigated, e.g. a COBRA (on 50 µm thin Polyimide film) [76]; it was designed for hole-multiplication and avalanche-ion blocking in gaseous photomultipliers (GPM). A more robust, Thick-COBRA (THCOBRA) [55] (Fig. 5d), e.g. on FR4 substrate, has expanded dimensions (similar to a THGEM, Figure 5b), with different surface-pattern configurations.

The THCOBRA operates in an MHSP-like two-stage avalanche mode. MHSP detectors yielded relatively large amplification factors, ~5x$10^4$ in Xe and ~5x$10^3$ in Ar gas media at 1 bar [77]. A more recent COBRA-like electrode was patterned on 125µm thick Polyimide film [77].
It is proposed here to detect both the ionization electrons and the primary scintillation with a MHSP or a COBRA-like electrode CsI-coated underneath, immersed in the noble liquid (L-MHSP; L-COBRA, L-THCOBRA), in a configuration similar to the one discussed in a gas medium [75].

The proposed detector is depicted in Figure 13. Here, the L-MHSP is immersed in the noble liquid, with the strips pointing up towards the photosensor array. Ionization electrons and scintillation-induced photoelectrons from CsI are collected effectively into the holes, under proper electric field configurations, similar to the situation and results described in 2.2. They are directed towards the anode strips; above the EL threshold, they induce VUV photons; additional photons are emitted above the avalanche threshold.

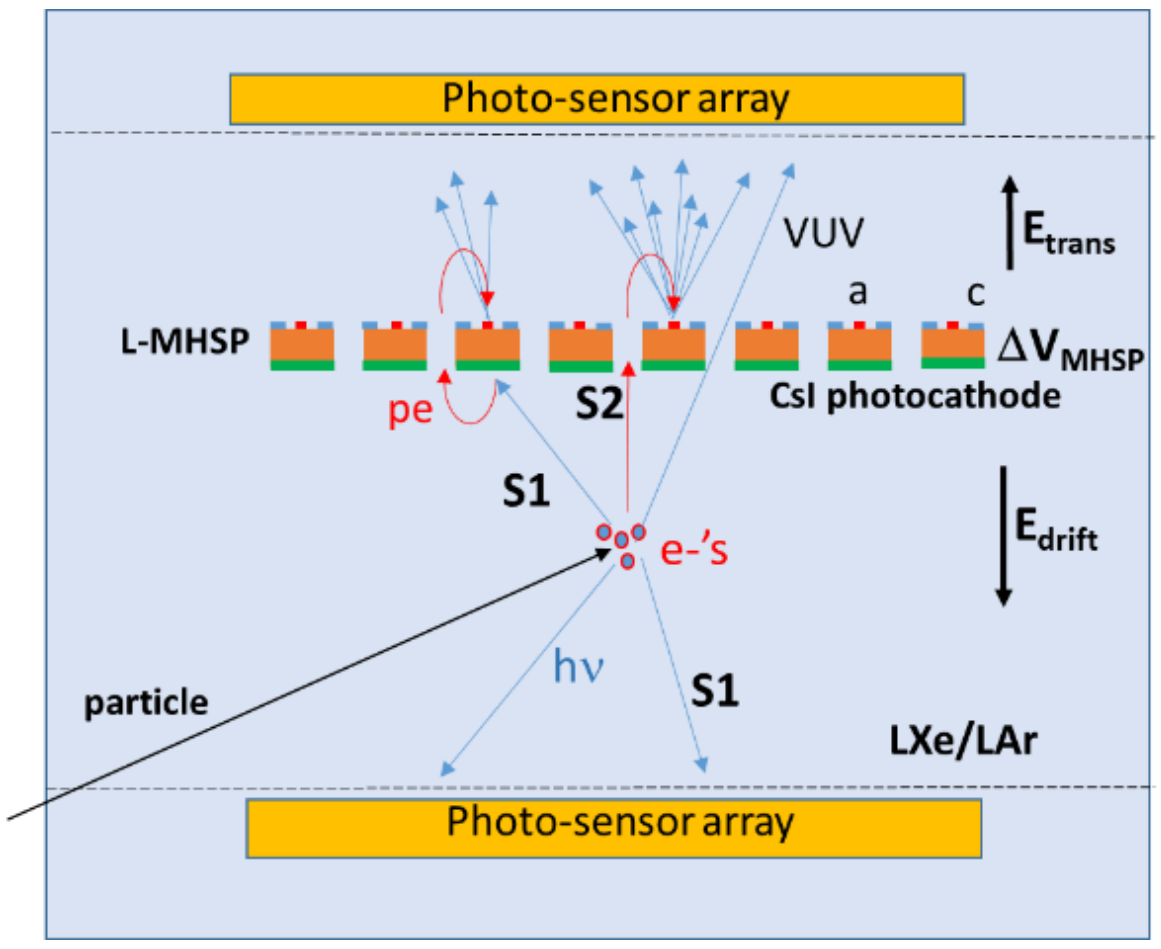

*Figure 13. A single-phase TPC with a L-MHSP (or L-COBRA) coated underneath with a CsI photocathode. Both ionization electrons and UV-induced photoelectrons from CsI are collected into the L-MHSP holes; they drift towards the anode strips, under proper electric field set across L-MHSP and potentials set on the anode "a" and cathode "c" strips. VUV photons emitted by EL + small avalanche near the anode strips, are detected (as well as a very small S1 photon fraction traversing the L-MHSP holes), by the top photo-sensors. Another fraction of S1 photons are either detected by bottom photo-sensors (as shown), or reflected by a mirror-cathode to the CsI photocathode (similar to Figure 7b). An optional wire mesh may be set under the L-MHSP, to enhance the pe extraction efficiency from CsI*

In this configuration, contrary to immersed wire arrays [31, 32, 33] and L-MSP configurations discussed in 3.1, all emitted VUV photons are viewed by the photosensors located nearby. Charge gain and EL yields in noble liquids were not yet measured with L-MHSP like structures; however, large charge gains are reported in a MHSP in various noble gases at pressures reaching several bar [77]; e.g. ~5x$10^3$ and ~5x$10^4$ at atmospheric pressure Ar and Xe, respectively. Photon yields (of the combined EL + avalanche) of~7x$10^4$ photons/electron were reported for an MHSP operated in Xe gas [54].

It should be also noted that, while MSP- and MHSP-based gas-avalanche detectors were designed for high-resolution localization, one could relax here on the hole-diameter and strip-pitch. The focus should rather be on optimization of the electron collection into the holes, extraction field at the photocathode surface ($QE_{eff}$) and on the patterned-electrode configuration – to maximize the total VUV-photoyield.

Unlike the L-MSP discussed in 3.1, the L-MHSP substrate does not have to be transparent – which should ease manufacturing of larger-area devices. Last, but not least, larger amplification factors may be reached by cascading two similar L-MHSP-like elements, both with photocathodes underneath (similar to the originally proposed LHM concept [39].

### 3.4. Single-phase detectors with Nano-structure coated electrodes

Electroluminescence and some avalanche gain in noble liquids might be reached at high electric fields set at Nano-structured surfaces. These have growing scientific and practical interests and applications; they are available, generally on small surfaces, in laboratory and industry, produced by various techniques; some examples are given in [79-82]. Such patterned surfaces of different kinds, including nanowires, rods, pyramids etc, are in use for solar cells, biosensors, photonics and more. Some examples of patterned surfaces are shown in Figure 14.

One could therefore imagine single-phase noble-liquid detectors, with immersed EL/charge multipliers, with perforated electrodes having nano-structures coating their top surface; optionally, they could be coated underneath with photocathodes, for VUV-photon sensing. For example, such single-element or cascaded devices could replace the MHSP-like configuration discussed in 3.2.

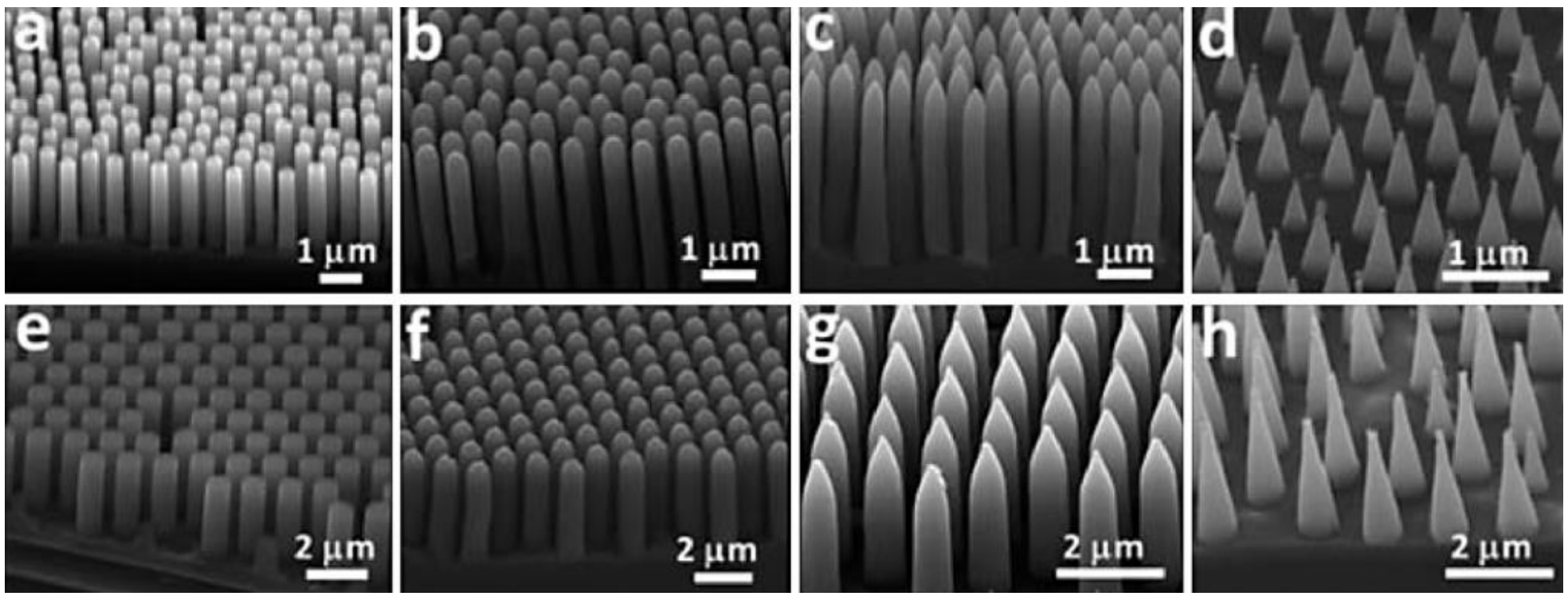

*Figure 14. SEM images of highly regular (a and e) nanopillar, (b and f) nanorod, (c and g) nanopencil, and (d and h) nanocone Si arrays, produced by wet-etching. (Reproduced from Ref. 81 with permission from The Royal Society of Chemistry).*

The concept is illustrated in Figure 15; electrons and scintillation-induced photoelectrons focused into the holes are collected at the electrode's top surface; they induce EL and most probably, some charge avalanche under the high local fields at the surface "tips".

While this concept sounds simple, the top surface should cover only the area around the holes; it has to be made homogenous, to avoid "hot spots" at local high fields and to assure good pulse-height resolution through photons detected by the photo-sensors. While current techniques might not yet be adequate for manufacturing such surfaces with the right materials, namely substrates and structures - over large areas, the field is growing rapidly. E.g. advanced 3D printing techniques already permit the fabrication of THGEM electrodes [83]; higher-resolution 3D printing techniques of polymer-made perforated electrodes, with metalized nano-structured surfaces, might become available in the near future, e.g. formed with amorphous Si on various substrate materials as reported in [79].

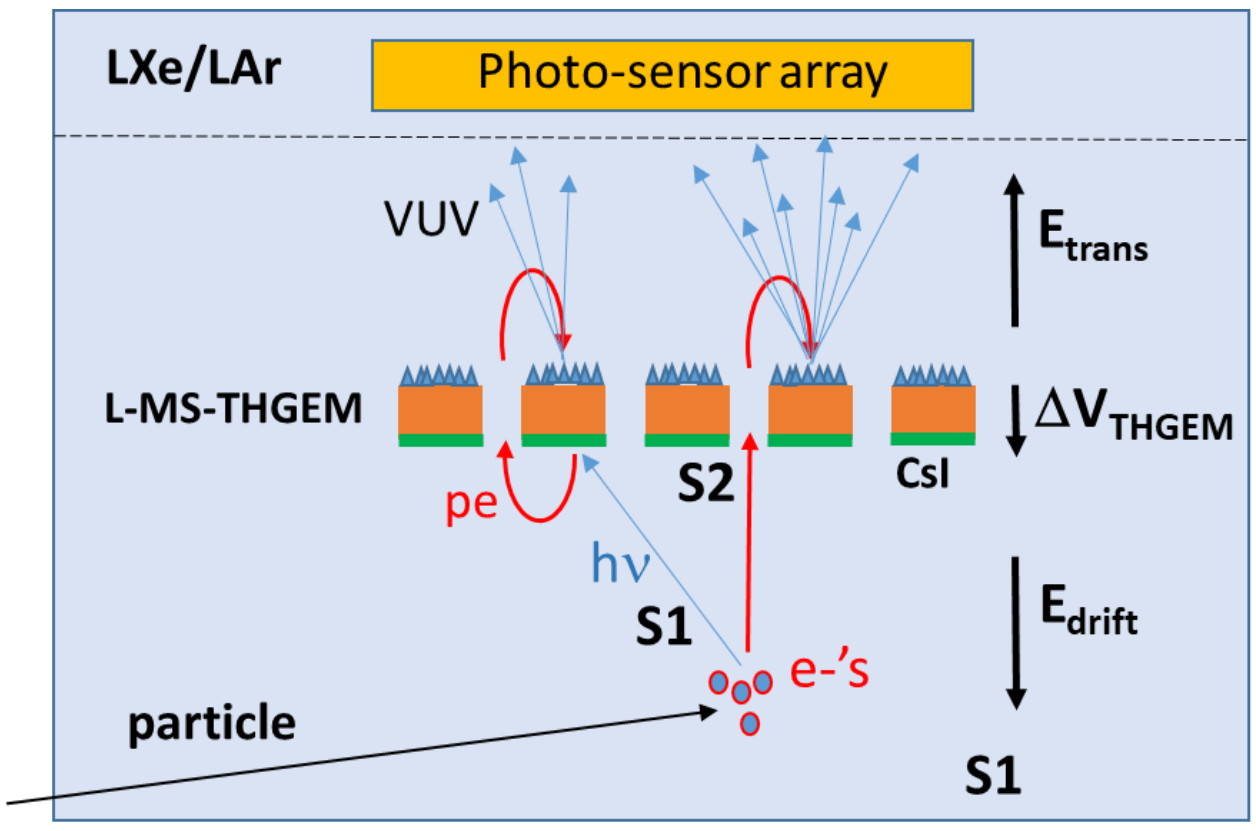

*Figure 15. A single-phase TPC with a Liquid nano-structured THGEM multiplier (L-MS-THGEM) coated underneath with a CsI photocathode. Similar to the L-MHSP (of Figure 13), both S2 ionization electrons and UV-induced photoelectrons from CsI are collected into the holes; they drift through the THGEM electrode towards the nano-structured top surface. VUV photons emitted by EL + small avalanche at the vicinity of the "anode tips", are detected by near-by photo-sensors. Another fraction of S1 photons are either detected by bottom photo-sensors or reflected by a mirror-cathode to the photocathode (as shown in Figure 7b). An optional wire mesh may be set under the L-NS-THGEM, to enhance the pe extraction efficiency from CsI.*

It should be added that such micro- and nano-structured surfaces, and others, e.g. porous ones formed by surface etching, could be coated with photocathodes (e.g. CsI). This would enhance the QE due to the larger emitting surface [84]; at very high fields, it would enhance photoemission due to reduction of the work function – as demonstrated in [85].

## 4. Summary and discussion

We presented here a collection of ideas, devising some currently-investigated and other novel concepts of electron and VUV-photon sensing in noble liquid detectors. The aim is to improve current single- and dual-phase detector techniques, with a goal of reaching improved sensitivity and resolutions.

The article was written with the intension of triggering an interest among young researchers in the field of Radiation Detection Physics.

The concepts proposed make use of known gas-avalanche detector-electrode types: GEM, THGEM, MSP of different strip configurations, and combined hole-and-strip MHSP and COBRA-like ones. According to the concepts discussed in this work, the detector electrodes operate within the noble liquid only or in both gas and liquid phases.

One of the goals is to "bypass" the technical issues related to the geometry and stability of the liquid-gas interface in dual-phase TPCs; these may cause construction and operation problems in future large-volume detectors, equipped with large-area grids. The bubble-assisted LHM (2.1), being indeed a "local" dual-phase detector, is sensitive to both ionization electrons and VUV scintillation photons; it was proposed to solve the interface-instability issues, with EL occurring in well-defined gas bubble trapped under an immersed perforated electrode, coated with a photocathode.

So far, we have encountered difficulties in effectively transferring electrons through the perforated electrode, via the liquid-gas interface, into the bubble – thus affecting both energy resolution and PDE. It is known, that electrons in noble liquid form a cloud of polarized molecules, leading to an effective potential well of a 0.69eV depth [3]. Thus, to pass from liquid-to-gas, an electron needs either to tunnel through this barrier or to be at the high-energy tail of a thermal distribution. Like in the dual-phase TPC, electrons are "pushed" against the flat interface. However, according to our current hypothesis [43, 47], in the LHM the field has a component tangent to the liquid-gas interface, that makes the electrons "glide" sideways over the interface; it causes them to either tunnel through the barrier or reach the bottom of the electrode's hole and get collected and lost. This mechanism depends on the geometry, electric fields and interface shape. Preliminary modeling and experimental efforts have been undertaken [49] to understand the processes involved and to hopefully devise methods for enhancing electron transfer into the bubble; e.g. by adequate shaping of the electric field at the liquid-to-bubble interface.

The proposed novel dual-phase "bubble-less" LHM detector concept aims at bypassing the interface instabilities in a different way. It relies on a two-stage process: an immersed CsI-coated perforated electrode collects ionization electrons and scintillation-induced photoelectrons; they are transferred effectively, through the holes, from liquid-to-gas into a second perforated electrode. It should be remembered that the immersed photocathode benefits from high effective quantum efficiency. Contrary to EL occurring along the parallel gas gap in standard dual-phase detectors, here the electrons induce EL- and additional avalanche-photons in small-volume holes of a GEM, THGEM or other electrodes, of well-defined geometry. MHSP- or COBRA-like electrodes, with additional strip amplification, would provide higher photon yields. The VUV photons are recorded with photo-sensors located above. The resulting S1 and S2 pulses are fast (in the few-ns or tens-of-ns range, according to the electrode geometry), in comparison with current parallel-gap detectors; the latter yield long EL-photon pulse trails. Single photons hitting the immersed CsI photocathode, result in intense VUV flashes in the gas-located electrode. Therefore, the resulting large signals will be considerably above the typical dark-current ones of the photo-sensors; this paves the way towards more economical photon-readout solutions. So far, the efficient transfer of electrons and photoelectrons through a THGEM perforated electrode immersed in LXe, has been demonstrated [50]; their extraction and focusing into the holes of the gas-located electrode and the resulting photoyield with various multiplier types, are the subject of current studies.

Some new proposed concepts of single-phase noble-liquid TPCs make use of microstrip plates (MSP), having thin strip anodes patterned in various geometries on solid substrates immersed in the liquid. The general idea is to detect and localize EL (and possibly some avalanche) VUV-photons induced near the strips, by ionization electrons and photoelectrons - with nearby photosensors.

In the simplest "practical" approach proposed, relevant to DM and neutrino-physics experiments, a MSP electrode with thin anode strips is deposited on a VUV-transparent substrate. Not being limited by a required high localization resolution, the preferred electrode would be of a VCC [66] configuration. Compared to a standard MSGC, with neigbouring anode and cathode strips, in the VCC the insulator substrate separates the anode strips from the back-plane cathode; this considerably increases the electrical rigidity, which permits applying higher potentials – expecting larger photoyields. The latter, are dictated by the electrode geometry and applied potentials - both affecting the field intensity at the vicinity of the anode strips. Few-nm thin Cr or Ni strips and the back-plane surface would transmit a significant fraction of the VUV photons to the photo-sensors. Preliminary simulations indicate that the EL threshold in LXe is located about 20 μm from the strip surface. The resulting photoyields of the various MSP configurations can be simulated in noble liquids, e.g. with similar approaches to those in noble gases [86].

The combined detection of ionization electrons and primary-scintillation photons can be performed by inserting a perforated electrode (e.g. GEM, THGEM), under the strip electrode, under-coated with a CsI (or other high-QE) VUV photocathode. Additionally, its top face can be coated with a reflector, for enhancing the detectable VUV-photon yied; it could be also coated with a WLS, which would permit using a glass substrate transmitting the visible-photons. Note however, that both solutions would somewhat affect the event's localization resolution. Both electrons and photoelectrons are efficiently collected into the holes and transferred to the strip electrode (as discussed above in 2.2). Note, that unlike single-photon detection with PMTs, SiPMs etc, here a single primary-scintillation photon would result in an intense secondary-photon flash. Thus, the usual photodetector dark currents are not expected to be an issue (e.g. misidentified as single-photon events). This should lead, for example, to lower detection thresholds in DM and other experiments.

MHSP-like and the COBRA-like perforated electrodes, coated underneath with VUV photocathodes, viewed from above by nearby photosensors, would most probably provide the *preferable solution* for single-phase noble-liquid TPCs. Unlike in a gas phase, EL and some charge multiplication of the collected electrons and photoelectrons will occur only at the anode-strip vicinity. The anode strips should be kept distant from the cathode patterns surrounding the holes, to permit stable operation at high applied potentials.

A more "futuristic" approach discussed here, relies on immersed multipliers consisting of perforated electrodes coated with homogenous nano-structured surfaces. Charge-induced EL (and possibly electron avalanche) are expected to occur at high local fields. Current and future technologies should permit the formation of such coated electrodes, with etched holes, on various substrate materials.

The novel concepts proposed here are based on optical readout. They require extensive investigations, preceeded by careful simulations of the expected photoyields. High experimentally-reachable yields would permit the use of economical optical-readout schemes. Among matters of concern necessitating experimental clarifications, are long-term stability, e.g. electrodes' charging up and dark currents induced e.g. by spontaneous electron (photon) emission, the eventual formation of bubbles under the electrodes etc. The enhanced photoemission properties in noble liquids were reported so far for CsI [46, 47]; the phenomenon was discussed as due to the favorable effect of the negative electron affinity in noble liquids [46] on the electron extraction. It would be worth investigating other VUV-sensitive photocathodes in the liquid phase, e.g. other alkali halides [88], diamond films [89] etc, that might present similar behavior. Note however that among known "stable" photocathodes, CsI has the highest QE value in vacuum at Xe emission wavelength; at Ar wavelength, CsI, CsBr and possibly boron-doped diamond might have similar QE values in vacuum [90].

It should be also taken into account that some of the detector concepts, e.g based on MSP-like multipliers on quartz and glass substrates, might be limited in size, due to technology limitations. In such cases, large detectors would require modular elements. Note that GEM and THGEM electrodes are produced over large area by the printed-circuits industry; large-area COBRA-like structures can most probably be produced in similar ways. An important recent development is the glass-made THGEM (G-THGEM), within the ARIADNE optical-TPC project [91]. So far, borofloat 33 glass and fused silica glass electrodes (the latter of higher radiopurity) were produced by abrasive formation of sub-mm holes. The electrode surfaces were coated by resistive indium tin oxide (ITO) resistive films; they can be patterned, according to the authors, by laser techniques – e.g. to form COBRA-like patterns. While current G-THGEM electrodes for ARIADNE reach sizes of 50x50cm$^2$, the technology can be extended to larger dimentions and to other glass types [91].

Thin anode strips and other metallic patterns are currently formed in industry, e.g. by inkjet [92] and photolithographic techniques [93]; the latter already permits the formation of few-micron thin strips of relatively large areas (up to 24"x 24") on a variety of substrate materials. CsI photocathodes are commonly deposited over large electrodes, e.g. for Ring Imaging Cherenkov (RICH) detectors [94]. Techniques of resistive DLC-coating over large-area detector electrodes, to reduce charging up effects, are well established [95]. Radiopurity of the electrode substrates should be of a major concern in rare-event (e.g. DM) experiments. E.g. Polyimide (Kapton) substrates of GEM and MHSP have relatively low radioactivity, and were also proposed for Cirlex-THGEM detectors in DM experiments [96]. A recent work [97] states that Kapton ELJ of DuPont has a single digit pg/g levels of $^{238}$U and $^{232}$Th, two-to-three orders of magnitude radiopure than commercial off-the-shelf ones. Moreover, copper-clad Kapton ELJ laminates have shown to be very radiopure: 110 and 89 μBq/kg of $^{238}$U and $^{232}$Th, respectively.
A broader discussion on this subject, as well as on other technical and economical issues, are beyond the scope of this work.

Unlike dual-phase detectors, the single-phase concepts proposed here may be deployed in face-to-face geometries (half the drift potential per given volume), as well as in horizontal (drift) configuration; the latter would be benefical for suppressing potential instabilities due to uncontrolled bubble formation within the liquid. Modular face-to-face elements (not necessarily of the same multiplication configurations) can be piled-up in some applications, to reduce the values of the applied drift potentials. They are expected to solve some technical problems in large-volume DM detectors, and reduce the influence of the photo-sensor's dark current – thus reducing detection thresholds. The lack of liquid-to-gas interface will avoid the usual S1-photon losses by total reflection, and issues related to delayed detection of a fraction of them – both affecting the energy resolution in current dual-phase detectors. The faster expected signals with some of the proposed patterned electrodes, compared to EL ones induced along a gas gap, offer the possibility of effective electron-counting (e.g. see [98]) analysis schemes discussed in [32]; these are expected, like with thin wire planes, to provide superior energy resolutions. It should be noted that the proposed wire-plane solution for ionization-electron detection [31, 32, 33] could be extended to scintillation-photon detection - if preceded with a perforated electrode coated with a photocathode underneath (similar to the concept discussed in Par.3.2).

Neutrino experiments, requiring very large single-phase detection volumes, could benefit from some (even modest) ionization-electron amplification via EL with some of the proposed concepts; it would result in lower detection thresholds, which might be of interest in some physics searches. It should be stressed though, that while EL and charge multiplication threshold fields exist for LXe [31] (shown in Figure 10), the data for LAr (employed in DM and large neutrino experiments) are yet controversial, as discussed in [99].
Out of the ideas presented here, we expect the most suitable concept for ionization-electron and scintillation-photon detection in large-volume noble-liquid detectors (with currently known technologies) to be a single-phase TPC with modular, robust, optically-recorded COBRA-like multipliers; their bottom face being coated with a VUV photocathode. cascaded COBRA-like multipliers, would provide larger amplification factors - as originally proposed for the single-phase LHM concept with immersed perforated electrodes [39]. However, the larger expected VUV-photon spread, induced by a two-stage EL amplification, may affect the localization properties.

In parallel to the ongoing research on the LHM-based concepts presented in Par. 2, simulations, experimental validation and scaling-up feasibility of the various newly-proposed ones are necessary. Some could be deployed at the circumference of the recently proposed Radial DM detectors: the Geiger Geometry TPCs (GG-TPC) [100] and Radial TPC (RTPC) [101]. Other applications, besides those discussed in this article, include Compton Cameras in astronomy [102] and medical imaging [103, 104] and in large-area detectors for fast-neutron & gamma imaging; e.g. the latter in the field of contraband detection [105].

## Acknowledgements

The R&D on the LHM and the dual-phase LHM projects, discussed here, has been performed at the Weizmann Institute. I would like to thank my Weizmann Institute and BGU collaborators, in particular Drs. Lior Arazi, Eran Erdal, Gonzalo Martínez Lema, Arindam Roy, & David Vartsky, for their contributions to the LHM-based projects and for their assistance in preparing this document. I'm particularly thankful to Gonzalo, for the field simulations in the MSP-like configurations. I'm indebted to my friends Profs. Alexey Buzulutskov (Budker Inst.) and Vitaly Chepel (Coimbra Univ.) for their useful remarks. This research activity has been carried out in the context of the DARWIN dark-matter and CERN-RD51 collaborations.